\documentclass[twocolumn,prb,showpacs,preprintnumbers,amsmath,amssymb]{revtex4}
\def\journal #1#2#3#4{#1 {\bf #2}, #3 (#4)}

\def\PRB{Phys.\ Rev.\ B}
\def\PRL{Phys.\ Rev.\ Lett.}

\def\JPCM{J.\ Phys.\ Cond.\ Mat.}

\def\JPSJ{J.\ Phys.\ Soc.\ Jpn.}

\usepackage{graphicx}
\usepackage{dcolumn}
\usepackage{bm}
\usepackage{amsmath}
\usepackage{times}
\usepackage[usenames]{color}

\newcommand{\means}[1]{\langle#1\rangle}

\usepackage{ulem}

\begin{document}

\title{Polar Charge Fluctuation and Superconductivity in Organic Conductor}
\author{Akihiko~Sekine$^{1 \ast}$, Joji~Nasu$^{1}$, and Sumio~Ishihara$^{1, 2}$} 
\affiliation{$^1$Department of Physics, Tohoku University, Sendai 980-8578, Japan}
\affiliation{$^2$Core Research for Evolutional Science and Technology (CREST), Sendai 980-8578, Japan}

\date{\today}
\begin{abstract}
Superconductivity and polar charge fluctuation are studied in an organic conductor with the dimer-molecule degree of freedom. The extended Hubbard models, where the intra-dimer electronic structure and the inter-dimer Coulomb interactions are taken into account, are analyzed by the random-phase approximation and the fluctuation-exchange approximation. Superconductivity appears in a vicinity of the charge-density wave (CDW) phase where the electronic charge distributions are polarized inside of dimers. The extended $s$-wave type paring is favored and its competitive relations with the superconductivity due to the spin fluctuation depends on the triangular lattice geometry. Comparison between two superconductivities realized near the polar and non-polar CDW phases are also presented. 
\end{abstract}


\maketitle
\narrowtext
%

Unconventional superconductivity is one of the central issues in modern solid state physics. Beyond the standard Bardeen-Cooper-Schrieffer theory, non-phononic mechanisms and anisotropic pairing symmetries have been proposed theoretically and examined experimentally in a wide variety of materials, such as copper oxides, iron pnictides, heavy fermion compounds and so on. Organic conductors are one of the families in which exotic superconductivities have been examined intensively and extensively since the discovery of superconductivity in the TMTSF salts.~\cite{Jerome80,Ardavan2012}

Recently much attention has been attracted in layered quasi two-dimensional $\kappa$-type BEDT-TTF salts, $\kappa$-(BEDT-TTF)$_2$X (X: a monovalent anion).~\cite{Kanoda2006,kanoda_chem} The maximum superconducting (SC) transition temperature ($T_{\rm c}$) of 13.2K is observed in $\kappa$-(BEDT-TTF)$_2$Cu[N(CN)$_2$]Cl under high pressure.~\cite{Williams91} In the conducting BEDT-TTF layers, two BEDT-TTF molecules are regarded as a molecular dimer which builds a two-dimensional triangular lattice. A geometrical anisotropy in the triangular lattice is adjustable by changing an anion $X$ and/or applying pressure. One hole-carrier per dimer occupies the lowest-unoccupied molecular-orbital in a dimer. When the intra-dimer Coulomb interaction is much stronger than the band width, the system is regarded as a Mott insulator, which is termed a so-called dimer-Mott insulator.~\cite{kino1,Kino1996} An antiferromagnetic ordered state as well as a spin-liquid state observed in a series of materials are consequences of the Mott insulating state.~\cite{shimizu,syamashita,myamashita} Since the SC phase appears at a vicinity of the magnetic phases,~\cite{Kanoda2006} a mechanism of the superconductivity has been often examined from the view point of the spin fluctuation.~\cite{kino2,Kondo1998,Kondo2001,kuroki,kondo,Watanabe2006} 

Recently, dielectric anomaly is reported in one of the $\kappa$-type salts, $\kappa$-(BEDT-TTF)$_2$Cu$_2$(CN)$_3$.~\cite{jawad} A cusp-like structure and strong frequency dependence are observed in the temperature dependence of the dielectric constant around 30K. One possible origin of this dielectric anomaly is attributed to the electric dipole moments inside of the BEDT-TTF molecular dimers, i.e. an electronic-charge distribution without the space-inversion symmetry in a dimer.~\cite{naka_kappa,hotta,li} Similar dielectric anomalies are also reported in $\kappa$-(BEDT-TTF)$_2$Cu[N(CN)$_2$]Cl~\cite{Lunkenheimer} and $\beta'$-(BEDT-TTF)$_2$ICl$_2$.~\cite{iguchi} These experimental results trigger reinvestigations of the electronic structure in the dimer-Mott insulating systems, and a mechanism of the superconductivity. 

In this paper, we study superconductivity in low-dimensional organic conductors, in particular, focus on a role of the polar charge fluctuation. We introduce the extended Hubbard model where, in contrast to the previous works, we take into account both the electronic degree of freedom inside of a dimer and the inter-dimer Coulomb interactions, which are essential for the polar charge order and fluctuation. Two types of the molecule configurations where two and four molecules are included in a unit cell are analyzed by the random-phase approximation (RPA) and the fluctuation-exchange (FLEX) methods. 
The SC phase appears near the charge-density wave (CDW) phase where the electronic charge distribution is polarized inside of dimers, as well as near the spin-density wave (SDW) phase. The extended $s$-wave symmetry pairing is favored near this CDW phase and is realized cooperatively with the $d_{xy}$-type pairing due to the spin fluctuation. Contribution of the polar charge fluctuation is remarkable around the so-called frustration point where the dimer molecules are arranged on an equilateral triangular lattice. A reentrant feature of the phase boundary is observed for the superconductivity induced by the polar charge fluctuation. 

\begin{figure}[t]
\includegraphics[width=0.8\columnwidth,clip]{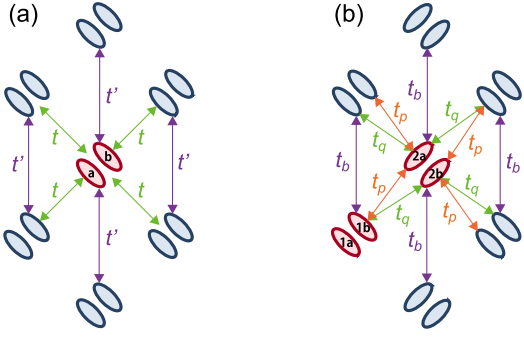}
\caption{(Color online) Schematic views of the two-band model in (a) and the four-band model in (b). The Coulomb interactions, $V$, and $V'$ in the two band model, and $V_b$, $V_p$, and $V_q$ in the four band model are introduced in the same ways as $t$ and $t'$, and $t_b$, $t_p$ and $t_q$, respectively. Ovals represent the BEDT-TTF molecules. 
}
\label{fig:lattice}
\end{figure}

We introduce the two types of the molecule configurations, termed the two-band model and the four-band model, where inequivalent two molecules and four molecules are introduced in a unit cell, respectively. Schematic views are shown in Fig.~\ref{fig:lattice}. First we present the Hamiltonian for the two-band model:
\begin{align}
{\cal H}=
&-\sum_{i \sigma} t_0
\left ( 
c_{i a \sigma}^\dagger c_{i b \sigma}+{\rm H.c.}
\right ) +U \sum_{i \gamma} n_{i \gamma \uparrow} n_{i \gamma \downarrow}\nonumber \\
&+U'\sum_{i } n_{i a} n_{i b}
-\sum_{\langle ij \rangle \gamma \gamma'\sigma}
t_{ij}^{\gamma \gamma'} \left ( 
c_{i \gamma \sigma}^\dagger c_{j \gamma' \sigma} +{\rm H.c.} \right )
\nonumber \\
&+\sum_{\langle ij \rangle \gamma \gamma'}
V_{ij}^{\gamma \gamma'} n_{i \gamma} n_{j \gamma'} , 
\label{eq:ham}
\end{align}
where $c_{i \alpha \sigma}$ is an electron annihilation operator at the $i$-th dimer with molecular orbital $\gamma(=a, b)$, and spin $\sigma(=\uparrow, \downarrow)$, and $n_{i \gamma}(\equiv \sum_{\sigma} n_{i \gamma \sigma}=\sum_{\sigma} c_{i \gamma \sigma}^\dagger c_{i \gamma \sigma})$ is the number operator. The first term represents the intra-dimer electron transfer, and the second and third terms represent the intra-molecule and inter-molecule electron-electron interactions inside of a dimer, respectively. The last two terms describe the inter-dimer electron transfers and electron-electron interactions, respectively. 
In the two-band model, the inter-dimer transfers and Coulomb interactions are assumed to be diagonal as $t_{ij}^{\gamma \gamma'}=\delta_{\gamma \gamma'} t_{ij} $ and $V_{ij}^{\gamma \gamma'}=\delta_{\gamma \gamma'} V_{ij} $, for simplicity. The geometrical anisotropy in a triangular lattice is represented by the two kinds of $t_{ij}$, $t, t'$, and those of $V_{ij}$, $V, V'$, as shown in Fig.~\ref{fig:lattice}(a). Spatial difference between the two molecules in the same dimer is neglected, for simplicity. The average electron number per dimer is fixed to be 3, corresponding to the averaged electron density in $\kappa$-type BEDT-TTF salts.
The interaction terms, i.e. the $U$, $U'$ and $V_{ij}^{\gamma \gamma'}$ terms in Eq.~(\ref{eq:ham}), are represented in a unified fashion as 
\begin{align}
{\cal H}_{\rm int}=\sum_{\bm k \bm p \bm q} 
\sum_{\alpha \beta \gamma \delta} \sum_{\sigma \sigma'}
[W_{{\bm q}}^{\sigma \sigma'}]_{(\alpha \beta)(\gamma \delta)} 
c_{\bm k+\bm q \alpha \sigma}^\dagger c_{\bm k \beta \sigma}
c_{\bm p \gamma \sigma'}^\dagger c_{\bm p+\bm q \delta \sigma'} ,  
\label{eq:hinter}
\end{align}
where $c_{\bm k \beta \sigma}$ is the Fourier transform of $c_{i \beta \sigma}$, and the interactions $[W_{ {\bm q}}^{\sigma \sigma'}]_{(\alpha \beta)(\gamma \delta)}$ are the $4\times 4$ matrices.

In the four band model, inequivalent two dimers in a unit cell are identified by an index $\xi(=1,2)$ and the annihilation operator introduced in Eq.~(\ref{eq:ham}) is replaced as $c_{i \alpha \sigma} \rightarrow c_{i \xi \alpha \sigma}$ where $(\xi, \alpha)=\{(1, a) (1,b) (2,a) (2,b)\} $. The intra-dimer electron transfer, the intra-molecule and inter-molecule electron-electron interactions inside of a dimer are the similar forms with the first three terms in Eq.~(\ref{eq:ham}), respectively. Three kinds of the inter-dimer electron transfers, $t_b, t_p, t_q$, and the inter-dimer Coulomb interactions, $V_b, V_p, V_q$, are introduced, as shown in Fig.~\ref{fig:lattice}(b). The effective interactions $[W_{ {\bm q}}^{\sigma \sigma'}]_{(\tilde{\alpha} \tilde{\beta})(\tilde{\gamma} \tilde{\delta})}$ with $\tilde \alpha \equiv (\xi, \alpha)$ introduced in Eq.~(\ref{eq:hinter}) are the $16\times 16$ matrices.

The multi-band Hubbard models with the long-range Coulomb interactions are analyzed by the RPA and FLEX methods.~\cite{Takimoto2004,Nakano2006,Esirgen1999,Kobayashi2004,Yoshimi11} We consider the so-called bubble-type and ladder-type diagrams where the Coulomb interactions are considered up to the infinite order of the perturbation. For the Coulomb interactions $U$ and $U'$, both the ladder- and bubble-type diagrams are adopted, and for the inter-dimer interaction $V_{ij}^{\gamma \gamma'}$, the bubble-type diagrams are only considered, for simplicity.~\cite{onari,kobayashi,tanaka}
The spin susceptibility $\hat \chi^s_{\bm q}(i \omega_n)$ and the charge susceptibility $\hat \chi^c_{\bm q}(i \omega_n)$ in the RPA scheme are given by 
\begin{align}
{\hat \chi}_{{\bm q}}^{s,c}(i \omega_n)=
\left [ 1 \mp {\hat \chi}^0_{{\bm q}}( i \omega_n)  {\hat U}^{s,c}_{\bm q} \right ]^{-1} 
{\hat \chi}^0_{{\bm q}}(i\omega_n) , 
\label{eq:chiRPA}
\end{align}
where $-$ and $+$ signs are for the spin and charge susceptibilities, respectively, ''hat'' implies the matrix, and ${\hat \chi}^0_{{\bm q}}(i\omega_n)$ is the bare susceptibility with the Matsubara frequency $\omega_n$ defined by 
\begin{align}
{\chi}^0_{ {\bm q}(\alpha \beta)(\gamma \delta)}(i\omega_n)&=\frac{T}{N} \sum_{{\bm k} \epsilon_m}
G^0_{{\bm k} \delta \alpha }(i \epsilon_m) \nonumber \\
&\hspace{1.4cm}\times G^0_{ {\bm q+\bm k}\beta \gamma}(i \omega_n+i \epsilon_m) . 
\label{eq:chi0}
\end{align}
We define ${\hat U}^s_{\bm q}={\hat W}_{\bm q}^{\uparrow \downarrow}-{\hat W}_{\bm q}^{\uparrow \uparrow}$, ${\hat U}^c_{\bm q}={\hat W}_{\bm q}^{\uparrow \downarrow}+{\hat W}_{\bm q}^{\uparrow \uparrow}$, and the bare Green's function ${\hat G}^0_{\bm k}(i \epsilon_n)$.  
In the FLEX method, the spin and charge susceptibilities are calculated in the equations given in Eqs.~(\ref{eq:chiRPA}) and (\ref{eq:chi0}) where the bare Green's functions are replaced by the full Green's functions defined by the Dyson equation, 
$
{\hat G}_{\bm k}(i \epsilon_n)^{-1}=
{\hat G}^0_{\bm k}(i \epsilon_n)^{-1}-{\hat \Sigma}_{\bm k}(i \epsilon_n)$. 
The self energy is given by 
\begin{align}
\Sigma_{\bm k (\alpha \beta)}(i \epsilon_n)=&\frac{T}{N} 
\sum_{\bm k' \epsilon_m \mu \nu} 
V_{\bm k-\bm k' (\alpha \mu)(\nu \beta) }(i \epsilon_n-i \epsilon_m) 
\nonumber \\
&\times
G_{\bm k' (\mu \nu)}(i \epsilon_m) , 
\end{align}
where we introduce the effective interaction: 
\begin{align}
{\hat V}_{\bm q }(i \omega_n)
=&\frac{3}{2}{\hat U}^s_{\bm q} {\hat \chi}^s_{\bm q}(i \omega_n) {\hat U}^s_{\bm q}
+\frac{1}{2} {\hat U}^c_{\bm q} {\hat \chi}^c_{\bm q}(i \omega_n) {\hat U}^c_{\bm q}
\nonumber \\
&-\frac{1}{4} \left ( {\hat U}^s_{\bm q}+{\hat U}^c_{\bm q} \right ) {\hat \chi}^0_{\bm q}(i \omega_n) 
\left ( {\hat U}^s_{\bm q}+{\hat U}^c_{\bm q} \right )  \nonumber \\
&+\frac{3}{2} {\hat U}^s_{\bm q}-\frac{1}{2} {\hat U}^c_{\bm q} . 
\end{align}

It is convenient to introduce the following static susceptibilities: the static spin susceptibility 
\begin{align}
 \chi_{\bm{q}}^{\rm spin}=\int_0^\beta d\tau\means{S^z_{\bm{q}}(\tau)S^z_{-\bm{q}}},
\label{eq:chis}
\end{align}
the static charge susceptibility 
\begin{align} 
 \chi_{\bm{q}}^{\rm charge}=\int_0^\beta d\tau\means{Q_{\bm{q}}(\tau)Q_{-\bm{q}}} , 
\label{eq:chic}
\end{align}  
and the static polarization susceptibility 
\begin{align}
 \chi_{\bm{q}}^{\rm polar}=\int_0^\beta d\tau\means{P_{\bm{q}}(\tau)P_{-\bm{q}}},
\label{eq:chip}
\end{align}
where $\mathcal{O}(\tau)=e^{\tau \mathcal{H}}\mathcal{O}e^{-\tau \mathcal{H}}$, and $S_{\bm q}^z$, $Q_{\bm{q}}$ and $P_{\bm{q}}$ are the Fourier transforms of the spin, charge and polarization operators, respectively. These are defined by $S_{i}^z = (1/2)\sum_{\alpha(=a,b) }(n_{i \alpha \uparrow}-n_{i \alpha \downarrow})$, $Q_i=(1/2)\sum_{\alpha(=a,b) \sigma} n_{i\alpha \sigma}$ and $P_i=(1/2)\sum_{\sigma} (n_{ia \sigma}-n_{ib \sigma})$ for the two-band model, and $S_{i}^z = (1/2)\sum_{\xi(=1,2) \alpha(=a,b) }(n_{i \xi \alpha \uparrow}-n_{i \xi \alpha \downarrow})$, $Q_i=(1/2)\sum_{\alpha(=a,b) \sigma} (n_{i1 \alpha \sigma}- n_{i 2 \alpha \sigma})$ and $P_i=(1/2)\sum_{\xi(=1,2 ) \sigma} (n_{i \xi a \sigma}-n_{i \xi b \sigma})$ for the four-band model.
The static susceptibilities are represented by the spin- and charge-susceptibility matrices previously introduced. The explicit representations, for example for the two band model, are given as 
\begin{align}
\chi_{\bm{q}}^{\rm spin}
=\frac{1}{2} \sum_{\gamma_1, \gamma_2(=a,b)} 
\chi_{\bm{q} (\gamma_1 \gamma_1)(\gamma_2 \gamma_2)}^s(0) , 
\end{align}
\begin{align}
 \chi_{\bm{q}}^{\rm charge}=\frac{1}{2} \sum_{\gamma_1, \gamma_2(=a,b)} 
\chi_{\bm{q} (\gamma_1 \gamma_1)(\gamma_2 \gamma_2)}^c(0) , 
\label{eq:chicg}
\end{align}
and 
\begin{align}
 \chi_{\bm{q}}^{\rm polar}=\frac{1}{2} \sum_{\gamma_1, \gamma_2(=a,b)} 
\chi_{\bm{q} (\gamma_1 \gamma_1)(\gamma_2 \gamma_2)}^c(0)
\epsilon_{\gamma_1 \gamma_2} , 
\label{eq:chipg}
\end{align}
where $\epsilon_{\gamma_1 \gamma_2}=1$ for $\gamma_1=\gamma_2$  and $-1$ for $\gamma_1 \ne \gamma_2$. The formulae for the four-band model are given in the similar ways. 
On the analogy of the definitions of the static-susceptibilities, we introduce the "bare" static charge-susceptibility $\chi_{\bm{q}}^{0\ \rm charge}$, and the ``bare'' static polarization susceptibility $\chi_{\bm{q}}^{0\ \rm polar}$. These are defined in Eqs.~(\ref{eq:chic}) and (\ref{eq:chip}), where the expectations $ \langle \cdots \rangle$ are replaced by those without the Coulomb interactions, and are calculated by Eqs.~(\ref{eq:chicg}) and (\ref{eq:chipg}), where ${\hat \chi}_{\bm{q}}^c(0)$ is replaced by ${\hat \chi}_{\bm{q}}^{0}(0)$ in the two-band model.

The SC transition temperature and the gap function are calculated by the Eliashberg equation. The linearized equation is given by 
\begin{align}
\lambda \Delta_{\bm k (\alpha \beta)}^{(l)}(i \epsilon_n)&=-\frac{T}{N} \sum_{\bm{k}'\epsilon_m  \mu \nu \mu' \nu'}
\Gamma_{{\bm k}-{\bm k}' (\alpha \mu)(\beta \nu)}^{(l)} (i \epsilon_n-i\epsilon_m)
\nonumber \\
 \times & G_{\bm{k}' (\mu \mu')}(i \epsilon_m) G_{-\bm{k}' (\nu \nu')}(-i \epsilon_m) 
\Delta_{\bm k' (\mu' \nu')}^{(l)}(i \epsilon_m) , 
\end{align}
where $l=s (t)$ for a spin singlet (triplet) paring. A parameter $\lambda$ is introduced and $T_c$ is determined by the condition $\lambda=1$. We define the interactions 
\begin{align}
{\hat \Gamma}_{\bm q}^{(s)}&=\frac{3}{2} {\hat U}^s_{\bm q} {\hat \chi}^s_{\bm q} {\hat U}^s_{\bm q} -\frac{1}{2} {\hat U}^c_{\bm q} {\hat \chi}^c_{\bm q} {\hat U}^c_{\bm q}
+\frac{1}{2} \left ( {\hat U}^{s}_{\bm q}+{\hat U}^c_{\bm q} \right ) , 
\label{eq:gammas}
\end{align}
and 
\begin{align}
{\hat \Gamma}_{\bm q}^{(t)}&=-\frac{1}{2} {\hat U}^s_{\bm q} {\hat \chi}^s_{\bm q} {\hat U}^s_{\bm q} -\frac{1}{2} {\hat U}^c_{\bm q} {\hat \chi}^c_{\bm q} {\hat U}^c_{\bm q}
+\frac{1}{2} \left ( -{\hat U}^s_{\bm q}+{\hat U}^c_{\bm q} \right ) .
\label{eq:gammac}
\end{align}
We take 32$\times$32 (64$\times$64) momentum-point meshes in the Brillouin zone and up to 8,192 (16,384) Matsubara frequencies in the numerical calculations by RPA (FLEX). 

\begin{figure}[t]
\includegraphics[width=\columnwidth,clip]{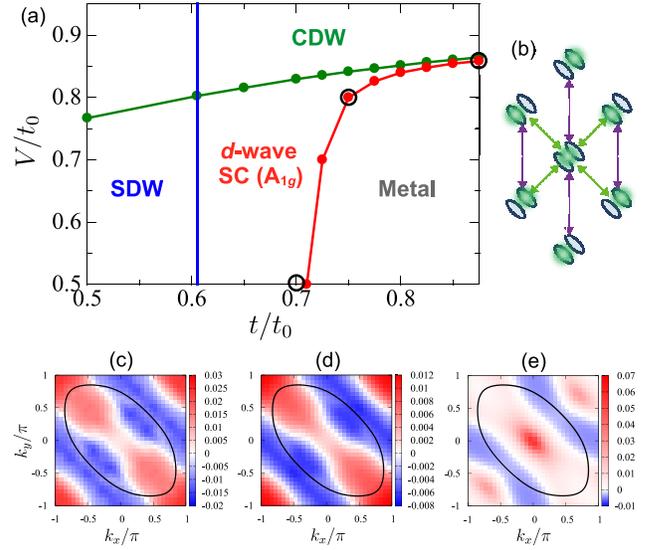}
\caption{(Color online)
(a) Phase diagram at $T/t_0=0.01$ for the two-band model calculated by RPA. (b) A schematic charge configuration in the polar CDW phases. Contour maps of the SC gap functions at $(t/t_0, V/t_0)=(0.7,0.5)$, $(0.75,0.8)$ and $(0.875,0.86)$, which are marked by circles in (a), are shown in (c), (d) and (e), respectively. Bold lines represent the Fermi surface. Parameter values are chosen to be $U/t_0=3$, $U'/t_0=2.1$, and $t'/t=V'/V=1.05$. 
}
\label{fig:twoRPA}
\end{figure}
First we show the numerical results in the two-band model. We take $t'/t=V'/V >1$ that correspond to the geometrical anisotropy in a triangular lattice for $\kappa$-(BEDT-TTF)$_2$Cu$_2$(CN)$_3$. The phase diagram at $T/t_0=0.01$ calculated by RPA is presented in Fig.~\ref{fig:twoRPA}(a). The topologically similar phase diagrams are reported in Refs.~\onlinecite{kobayashi} and \onlinecite{tanaka}, where one molecule without the dimer degree of freedom is introduced at each site in triangle and square lattices.  

The boundary for the SDW (CDW) phase is identified by the condition ${\rm det}[\bm{1} -(+) {\hat \chi}^0_{{\bm q}}(0)  {\hat U}^{s (c)}_{\bm q}]=0$. The SDW and CDW phases appear in small $t/t_0$ and small $V/t_0$ regions, respectively. The phase boundary for SDW does not depend on $V$, because the inter-dimer Coulomb interactions are not included in ${\hat U}^s_{\bm q}$. The momenta where the static spin susceptibility $\chi^{\rm spin}_{\bm q}$ takes the maxima are ${\bm q}=(0.75\pi, 0.375 \pi)$ and $(-0.75\pi, -0.375 \pi)$ at $(t/t_0,V/t_0)=(0.65,0.5)$.
To examine the CDW phase in more detail, we calculate the static charge and polar susceptibilities. At the boundary for the CDW phase, $\chi_{\bm q}^{\rm polar}$ diverges and $\chi_{\bm q}^{\rm charge}$ does not. That is, the charge disproportionation occurs inside of dimers and the electron number per dimer is equal for all dimers. 
A schematic charge configuration is shown in Fig.~\ref{fig:twoRPA}. The momenta where $\chi_{\bm q}^{\rm polar}$ takes the maxima are ${\bm q}=(2 \pi/3, 2\pi/3)$ and $(-2 \pi/3, -2\pi/3)$ at $(t/t_0, V/t_0)=(0.75, 0.8)$, implying a three-fold periodicity of the electric dipole moment and no macroscopic polarization. 

The SC phase boundary and the SC gap functions are shown in Fig.~\ref{fig:twoRPA}. The SC phase volume near the SDW phase is larger than that near the CDW phase. In the present lattice structure with the D$_{2h}$ point group, a spin-singlet SC-gap symmetry is classified by the A$_{1g}$ and B$_{1g}$ irreducible representations. The $d_{xy}$-type A$_{ 1g}$ SC gap is realized near the SDW phase at $(t/t_0, V/t_0)=(0.7,0.5)$ and $(0.75,0.8)$, and is gradually changed into the extended $s$-type A$_{1g}$ gap near the CDW phase.  This extended $s$-type gap is also obtained by the calculation where we set ${\hat U}^s {\hat \chi}^s {\hat U}^s=0$ in Eqs.~(\ref{eq:gammas}) and (\ref{eq:gammac}). It is concluded that the polar charge fluctuation induces the extended $s$-type SC state. 

\begin{figure}[t]
\includegraphics[width=\columnwidth,clip]{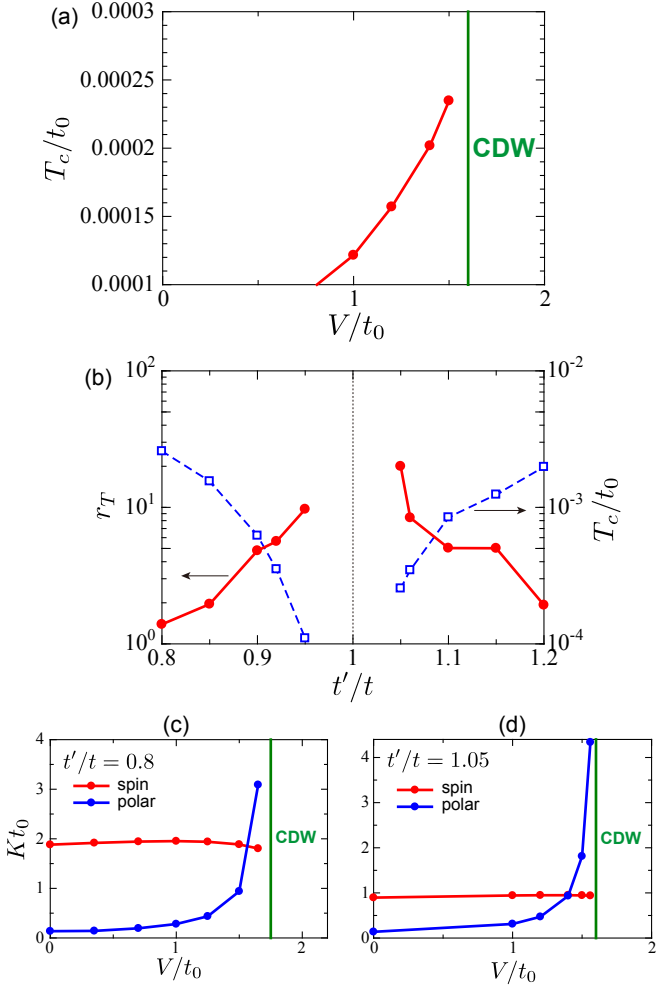}
\caption{(Color online)
(a) Superconducting transition temperatures near the CDW phase at $t'/t=V'/V=1.05$. (b) A ratio of the transition temperatures at $V/t_0=1.5$ to those at $V=0$, i.e. $r_T\equiv T_c(V/t_0=1.5)/T_c(V=0)$ as functions of the anisotropy in the electron transfer, $t'/t$. The transition temperatures at $V/t_0=1.5$ are also shown. The charge and spin susceptibilities at $T/t_0=0.0025$ for $t'/t=0.8$ and those for $t'/t=1.05$ are presented in (c) and (d), respectively. All data are obtained in the two-band model analyzed by FLEX. Other parameter values are chosen to be $U/t_0=4$, $U'/t_0=2.8$, and $t/t_0=0.5$. 
}
\label{fig:tcflex}
\end{figure}
The two-band model is further analyzed by the FLEX method. The CDW phase with the electric dipole moments inside of dimers, obtained by RPA [see Fig.~\ref{fig:twoRPA}], is reproduced, but the SDW phase is not realized in finite temperatures due to the Mermin-Wagner's theorem satisfied in FLEX.~\cite{kontani06} The SC transition temperatures near the CDW phase are shown in Fig.~\ref{fig:tcflex}(a). The SC transition temperatures are determined by fitting the numerical data of $\lambda$ calculated down to $0.002t_0$ by a function $\lambda=-a\log T+b$. The transition temperature monotonically increases, when the system approaches to the phase boundary. This is evidence that the SC state around this region is caused by the polar charge fluctuation. A numerical value of $T_c$ for $t'/t=1.05$ ($t'/t=0.8$) is estimated to be about 0.5K (5K) at $V/t_0=1.5$, when $t_0=0.2$eV is assumed. 
It is worth to note that, in comparison with the results by RPA [see Fig.~\ref{fig:twoRPA}(a)], the polar CDW phase and the SC phase near the CDW phase are realized until at least $t/t_0 =0.3$. This originates from a suppression of the SDW instability due to the Mermn-Wagner's theorem and is consistent with $\kappa$-type BEDT-TTF salts.~\cite{kanoda_chem} 

\begin{figure}[t]
\includegraphics[width=\columnwidth,clip]{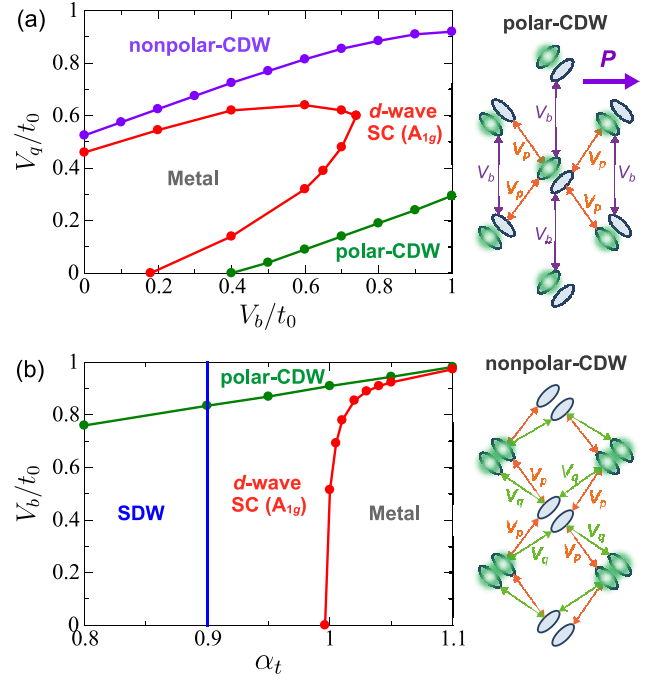}
\caption{(Color online)
(a) Phase diagram in the $V_b$-$V_q$ plane, and (b) that in the $\alpha_t$-$V_b$ plane at $T/t_0=0.1$ for the four-band model calculated by RPA. Schematic polar and non-polar CDW structures are also shown. We choose $t_b/t_0=0.5$, $t_p/t_0=0.35$, and $t_q/t_0=-0.13$ in (a), and $t_b/t_0=0.5\alpha_t$, $t_p/t_0=0.35\alpha_t$, and $t_q/t_0=-0.13\alpha_t$ in (b). Other parameter values are chosen to be $U/t_0=1.6$, $U'/t_0=1$, and $V_p/t_0=0.95$.
}
\label{fig:fourRPA}
\end{figure}
Effects of the triangular-lattice geometry on the SC transition temperature is examined by calculating a ratio of the transition temperatures at $V/t_0=1.5$ to that at $V=0$, i.e. $r_T\equiv T_c(V/t_0=1.5)/T_c(V=0)$ as functions of the anisotropy in the electron transfer, $t'/t$ [see Fig.~\ref{fig:tcflex}(b)]. We also plot the $t'/t$ dependence of $T_c$ at $V/t_0=1.5$, which is consistent with the results in Ref.~\onlinecite{kondo}.  It is clearly shown that $r_T$ takes its maximum around the so-called frustration point $t'/t=1$. That is, a relative contribution of the polar charge fluctuation to the spin fluctuation increases around $t'/t=1$. In Figs.~\ref{fig:tcflex}(c) and (d), the spin susceptibility and the polarization susceptibility at $T/t_0=0.0025$ are shown for $t'/t=0.8$ and 1.05, respectively. We define $K_s$ and $K_p$ are the maxima of $\chi_{\bm{q}}^{\rm spin}$ and $\chi_{\bm{q}}^{\rm polar}$, respectively, when the momenta $\bm q$ are varied. It is shown that $K_s$ at $t'/t=1.05$ is reduced from that in $t'/t=0.8$, on the other hand, $K_p$ are almost independent of $t'/t$. This is because the spin fluctuation is almost governed by $\chi^0_{\bm q}$ which is sensitive to the anisotropy of the electron transfer, $t'/t$, but the charge fluctuation is dominated by both the inter-molecule and inter-dimer Coulomb interactions. 

\begin{figure}[t]
\includegraphics[width=0.8\columnwidth,clip]{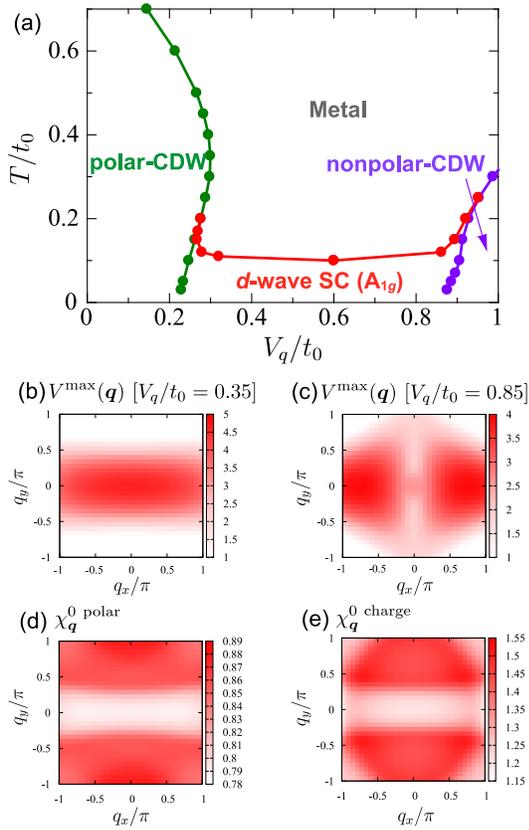}
\caption{(Color online)
(a) Finite temperature phase diagram for the four-band model calculated by RPA. Contour maps of $V^{\rm max}({\bm q})$, the largest eigenvalues of the Fourier transform of $V_{ij}^{\gamma \gamma'}$, are plotted in (b) for $V_q/t_0=0.35$ and in (c) for $V_q/t_0=0.85$. Contour maps of the ``bare'' static polar-susceptibility $\chi^{0\ \rm polar}_{\bm{q}}$ and the ``bare'' static charge-susceptibility $\chi^{0 \ \rm charge}_{\bm{q}}$ at $T/t_0=0.05$ are plotted in (d) and (e), respectively. The parameter values are chosen to be $U/t_0=1.6$, $U'/t_0=1$, $V_p/t_0=0.95$, $V_b/t_0=0.9$, $t_b/t_0=0.5$, $t_p/t_0=0.35$, and $t_q/t_0=-0.13$. 
}
\label{fig:reentrant}
\end{figure}
\begin{figure}[t]
\includegraphics[width=0.8\columnwidth,clip]{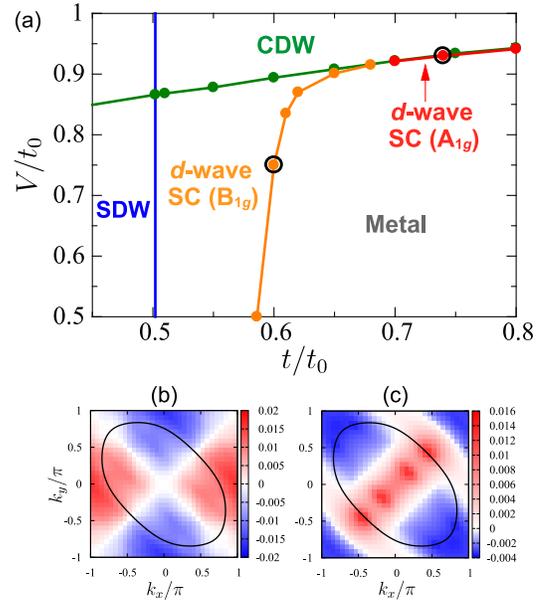}
\caption{(Color online)
(a) Phase diagram at $T/t_0=0.01$ for the two-band model for $t'/t=V'/V=0.9$ calculated by RPA. Contour maps of the SC gap function at $(t/t_0,V/t_0)=(0.6, 0.73)$, and $(0.74, 0.93)$, which are marked by circles in (a), are shown in (b) and (c), respectively. Bold lines represent the Fermi surface. The parameter values are chosen to be $U/t_0=2.4$, and $U'/t_0=1.6$.
}
\label{fig:b1g}
\end{figure}
The SC state induced by the polar charge fluctuation is studied in the realistic four-band model. In Fig.~\ref{fig:fourRPA}(a), the phase diagram in the four-band model calculated by RPA at $T/t_0=0.1$ is presented in the $V_b$-$V_q$ plane. Numerical values for the transfer integrals are chosen by the extended H\"uckel tight-binding band calculations.~\cite{Komatsu96} Two types of the CDW phases appear; the non-polar CDW for large $V_q$ where electron number per dimer are different in two inequivalent dimers, and the polar CDW for large $V_b$ where the polar charge distribution is realized inside of dimers. Schematic charge configurations are shown in Fig.~\ref{fig:fourRPA}. The SC phases are observed near both the phase boundaries for the polar and non-polar CDW phases. In Fig.~\ref{fig:fourRPA}(b), we plot the phase diagram in the $\alpha_t$-$V_b$ plane, where the inter-dimer transfers are given by a parameter $\alpha_t$ as $t_b/t_0=0.5\alpha_t$, $t_p/t_0=0.35\alpha_t$, and $t_q/t_0=-0.13\alpha_t$. This phase diagram qualitatively reproduces the results in the two-band model (see Fig.~\ref{fig:twoRPA}). We confirm that the SC phase induced by the polar charge fluctuation does not depend on a detailed lattice structure. 

Finite temperature phase diagram near the polar and non-polar CDW phases calculated by RPA is presented in Fig.~\ref{fig:reentrant} as a function of $V_q$. A reentrant feature is observed in the phase boundary between the polar CDW phase and the metallic or SC phase, but not in the boundary for the non-polar CDW phase. This is interpreted by the competitive momentum dependences of the charge susceptibility and the effective interaction as follows. In the RPA scheme, the momentum dependences of the susceptibilities are governed by those of the bare susceptibilities and the long-range Coulomb interactions. Contour maps of $V^{\rm max}(\bm{q})$, which is defined as the largest eigenvalues of the Fourier transforms of the inter-dimer interaction $V_{ij}^{\gamma \gamma'}$, are calculated. Results near the polar CDW phase ($V_q/t_0=0.35$) and the non-polar CDW phase ($V_q/t_0=0.85$) are plotted in Figs.~\ref{fig:reentrant}(b) and (c), respectively. We also show the contour maps of the ``bare'' static polar- and charge-susceptibilities, $\chi_{\bm{q}}^{0 \ \rm polar}$ and $\chi_{\bm{q}}^{0 \ \rm charge}$, in Fig.~\ref{fig:reentrant}(d) and (e), respectively. It is noticed that momentum dependences of $\chi_{\bm{q}}^{0\ \rm polar}$ and $V^{\rm max}({\bm q})$ at $V_q=0.35$ [see Figs~\ref{fig:reentrant}(b) and (d)] are competitive with each other; $\chi_{\bm{q}}^{0\ \rm polar}$ ($V^{\rm max}$) takes its maxima (minima) along $q_y=0$. This fact brings about a competition between the susceptibility and the inter-dimer interaction, and suppresses the instability toward the polar-CDW at low temperature. On the other hand, such competitive momentum dependences are weak for $\chi_{\bm{q}}^{0\ \rm charge}$ and $V^{\rm max}({\bm q})$ near the non-polar CDW phase [see Figs~\ref{fig:reentrant}(c) and (e)]. A similar reentrant feature and the competitions are suggested in Refs.~\onlinecite{kobayashi} and \onlinecite{Merino2001}.

Finally, we mention the SC properties in the two-band model with $t'/t<1$, that corresponds to $\kappa$-(BEDT-TTF)$_2$X for X=Cu(NCS)$_2$ and Cu[N(CN)$_2$]Cl. Phase diagram and gap functions calculated by RPA are presented in Fig.~\ref{fig:b1g}. In contrast to the results in Fig.~\ref{fig:twoRPA} where we take $t'/t>1$, the SC phase with the $d_{x^2-y^2}$-type B$_{1g}$ symmetry gap appears near the SDW phase. This change in the gap symmetry by changing $t'/t$ is consistent with the previous results.~\cite{kuroki,kobayashi} The extended $s$-type A$_{1g}$ gap is also realized near the CDW phase. As a consequence of the competition between the two SC phases with different symmetry gaps, a phase volume of the extended $s$-type SC is shrunk. We propose that the SC state induced by the polar charge fluctuation is favored in the case of $t'/t>1$. 

In summary, we study the superconductivity in the low-dimensional organic conductor where electronic structure inside of dimer is taken into account. We focus on roles of the polar charge fluctuation on the superconductivity. The two-types of the extended Hubbard models are analyzed by the RPA and FLEX methods. We find that the extended $s$-type SC state is realized near the CDW phase where the charge distributions inside of dimers are polarized. A monotonic increase of $T_c$ when the system approaches to the polar CDW phase is direct evidence of the SC state induced by the polar charge fluctuation. The obtained $T_c$ values are comparable to those in the $\kappa$-(BEDT-TTF)$_2$X systems. Competitive relationship between this SC state and that induced by the spin fluctuation depends on the geometrical anisotropy in a triangle lattice; two SC states are cooperative (competitive)  with each other in $t'/t>1$ ($t'/t<1$). The reentrant structures emerge in the boundary between the SC state and the polar-CDW state. The present study opens a new direction for research of the superconductivity and ferroelectricity. 

\begin{acknowledgments}
Authors would like to thank 
M.~Naka, S. Yamazaki, H.~Takashima, and T.~Watanabe for their valuable discussions. This work was supported by KAKENHI from MEXT and Tohoku University ``Evolution'' program. JN is supported by the global COE program ``Weaving Science Web beyond Particle-Matter Hierarchy'' of MEXT, Japan. Parts of the numerical calculations are performed in the supercomputing systems in ISSP, the University of Tokyo.
\end{acknowledgments}

*Present address: Institute for Materials Research, Tohoku University, Sendai 980-8577, Japan.

\vfill
\eject
\end{document}